\documentclass[a4paper,11pt]{article}
\pdfoutput=1 

\usepackage{jinstpub} 

\title{\boldmath Regular and Chaotic Motion Domains in the Channeling Electron's Phase Space and Mean Level Density for Its Transverse Motion Energy}

\author[a,b]{N.F. Shul'ga,}
\author[c,1]{V.V. Syshchenko,\note{Corresponding author.}}
\author[c]{A.I. Tarnovsky,}
\author[c]{V.I. Dronik}
\author[d]{and A.Yu. Isupov}

\affiliation[a]{Akhiezer Institute for Theoretical Physics of the NSC ``KIPT'',\\Akademicheskaya Street, 1, Kharkov 61108, Ukraine}
\affiliation[b]{V.N. Karazin National University,\\Svobody Square, 4, Kharkov 61022, Ukraine}
\affiliation[c]{Belgorod State University,\\Pobedy Street, 85, Belgorod 308015, Russian Federation}
\affiliation[d]{Laboratory of High Energy Physics (LHEP), Joint Institute for Nuclear Research (JINR),\\Dubna 141980, Russian Federation}

\emailAdd{syshch@yandex.ru}

\abstract{The motion of charged particles in a crystal in the axial channeling regime can be both regular and chaotic. The chaos in quantum case manifests itself in the statistical properties of the energy levels set. These properties have been studied previously for the electrons channeling along [110] direction of the silicon crystal, in the case when the classical motion was completely chaotic. The case of channeling along [100] direction is of special interest because the classical motion here can be both regular and chaotic for the same energy depending on the initial conditions. The semiclassical energy level density (as well as its part that corresponds to the regular motion domains in the phase space) is computed for the 10 GeV channeling electrons and positrons. It is demonstrated that the level spacing distribution for both electrons and positrons can be better described by Berry--Robnik distribution than by both Wigner (completely chaotic case) or Poisson (completely regular case) distributions.}


\arxivnumber{1909.13238} 

\proceeding{XIII$^{\text{th}}$ International Symposium <<Radiation from Relativistic Electrons in Periodic Structures>>\\
  September 16-20, 2019\\
  Belgorod, Russian Federation}

\begin{document}
\maketitle
\flushbottom

\section{Introduction}
\label{sec:intro}

When a fast charged particle is incident on a crystal at a small
angle to any crystallographic axis densely packed with atoms, it can
perform the finite motion in the transverse plane. This motion is
known as the axial channeling \cite{AhSh, AhSh2, Ugg}. The particle
motion in this case could be described with a good
accuracy as the one in the continuous potential of the atomic string.
During motion in this potential the longitudinal particle momentum
$p_\parallel$ is conserved, so the motion description is reduced
to two-dimensional problem of motion in the transversal plane. From the viewpoint of the dynamical systems theory, the channeling
 particle's motion could be either
regular or chaotic. The quantum chaos theory \cite{9,Stockmann,Reichl} predicts qualitative differences for these alternatives. 

The manifestations of chaos in quantum systems are found, first of all, in the statistical properties of their energy spectra. The quantum chaos theory predicts (see, e.g., \cite{9,Stockmann,Reichl}) that the energy levels nearest-neighbor distribution of the chaotic system obeys Wigner distribution
\begin{equation}
    \label{Wigner}
   p(s) = (\pi\rho^2 s/2) \exp (-\pi\rho^2 s^2/4) 
\end{equation}
(where $s$ is the distances between consequent energy levels, $\rho$ is the mean level density on the energy range under consideration) while the regular system --- the exponential one (frequently referred as Poisson distribution)
\begin{equation}
    \label{Poisson}
   p(s) = \rho\exp  (-\rho s) \,.
\end{equation}
The level statistics for the electrons channeling near $[110]$ direction in silicon crystal has been studied in \cite{Pov.2015, NIMB.2016}. In that case each pair of the closest parallel atomic strings forms the two-well potential (see, e.g., \cite{AhSh2}), in which the motion above the saddle point is chaotic for the major part of the initial conditions. As a result, the level spacing statistics in that case is well described by Wigner distribution (\ref{Wigner}).

The aim of the present paper is to study the level spacing statistics in the case of co-existence of regular and chaotic motion domains. Such situation takes the place when the electron channels near $[100]$ direction of silicon crystal. The level spacing statistics is described in this case by Berry--Robnik distribution \cite{Berry}
\begin{equation}\label{Berry.Rob}
p(s) = \frac{1}{\rho} \exp \left( -\rho_1 s \right)
\left\{ \rho_1^2 \mathrm{erfc}\, \left( \frac{\sqrt{\pi}}{2} \rho_2 s \right) + \left( 2\rho_1 \rho_2 + \frac{\pi}{2} \rho_2^3 s \right) \exp \left( -\frac{\pi}{4} \rho_2^2 s^2 \right) \right\},
\end{equation}
where 
\begin{equation}\label{erfc}
\mathrm{erfc}\, (x) = \frac{2}{\sqrt{\pi}} \int_x^\infty e^{-t^2} dt = 1 - \mathrm{erf}\, (x) \,.
\end{equation}
 It is presupposed that the regular motion domains and (single) chaotic motion domain generate independent level sequences with the densities $\rho_1$ and $\rho_2$  ($\rho = \rho_1 + \rho_2$) respectively. Our goal is to calculate the relative contribution of the regular and chaotic motion domains into the mean level density.

\section{Method and potential wells}
\label{sec:method}

\begin{figure}[htbp]
\vspace*{-2mm}\centering
\includegraphics[width=0.45\textwidth]{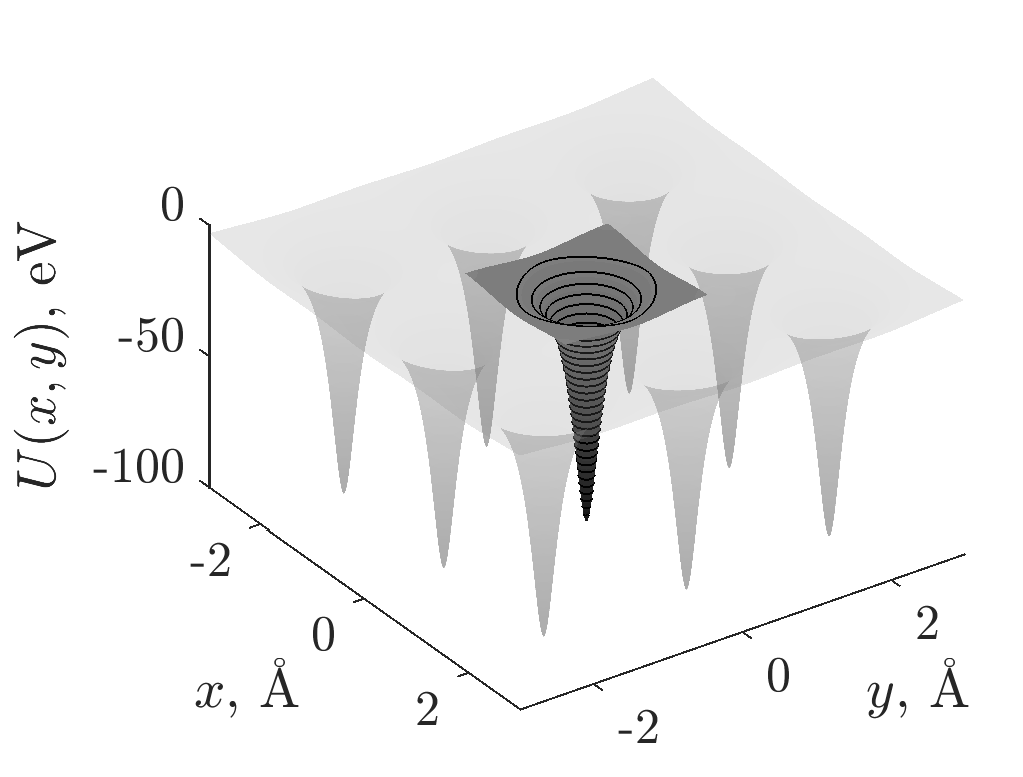}  \ 
\includegraphics[width=0.45\textwidth]{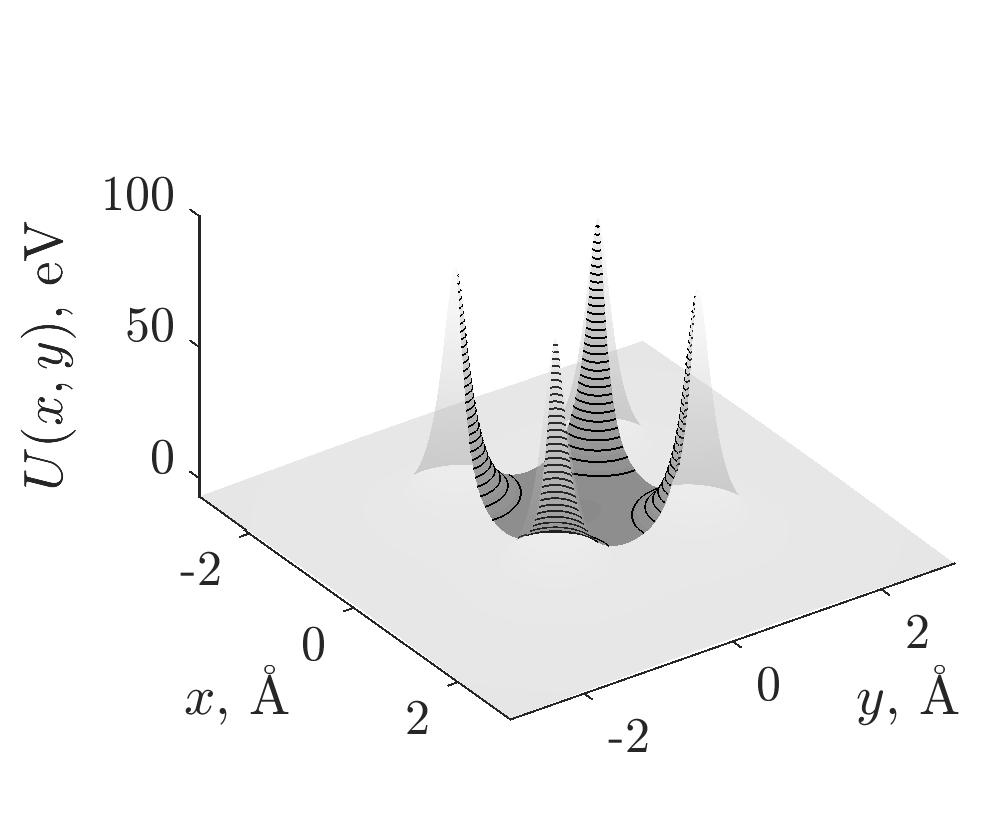} 
\vspace*{-2mm}\caption{\label{potentials} Potentials (\ref{U.el}) (left) and (\ref{U.pos}) (right).}
\end{figure}

The electron transversal motion in the atomic string continuous potential is described by the two-dimensional Schr\"odinger equation with Hamiltonian
\begin{equation}
    \label{Hamiltonian}
    \hat{H} = - (c^2\hbar^2 / 2 E_\parallel)  \left[ (\partial^2/\partial x^2) + (\partial^2/\partial y^2) \right] + U(x, y) \,,
\end{equation}
where the value $E_{\parallel} / c^2$ (here $E_{\parallel} = (m^2
c^4 + p_{\parallel}^2 c^2)^{1/2}$) plays the role of the particle mass
\cite{AhSh}. The Hamiltonian eigenvalues $E_\perp$ are found numerically using the so-called spectral method \cite{3, Dabagov3, NIMB.2013}. 
Here we consider the particle's motion near direction of the
atomic string $[100]$ of the Si crystal. The continuous potential
could be represented by the modified Lindhard potential
\cite{AhSh}
\begin{equation}
    \label{U.1}
    U^{(1)} (x, y) = - U_0 \ln \left[ 1 + \beta R^2 / (x^2 + y^2 + \alpha R^2) \right] \,,
\end{equation}
where $U_0 = 66.6$~eV, $\alpha = 0.48$, $\beta = 1.5$, $R =
0.194$~\AA~(Thomas--Fermi radius). These strings form in the plane
$(100)$ the square lattice with the period $a \approx 1.92$ \AA .  The additional contributions from the eight closest neighboring strings lead to
the following potential energy of the channeling electron (figure \ref{potentials}, left):
\begin{equation}
    \label{U.el}
    U^{(-)} (x, y) = \sum_{i=-1}^1 \sum_{j=-1}^1 U^{(1)} (x-ia, y-ja)  \,.
\end{equation}
Axial channeling of positrons near $[100]$ direction is possible in the 
small potential pit near the center of the square
cell with repulsive potentials $-U^{(1)}$ in the corners of the
square (figure \ref{potentials}, right):
\begin{equation}
    \label{U.pos}
    U^{(+)} (x, y) = -U^{(1)} (x-a/2, y-a/2 )
    -U^{(1)} (x-a/2, y+a/2) -
\end{equation}
\begin{equation*}
    -U^{(1)} (x+a/2, y-a/2)
    -U^{(1)} (x+a/2, y+a/2) - 7.9589\ \mathrm{eV} \,,
\end{equation*}
where the constant is chosen to achieve
zero potential in the center of the cell. The spectral method for the channeling electrons and positrons near $[100]$ direction had been tested in \cite{group} for small $E_\parallel$ values, when the total number of energy levels in the potential well is small. Here we put $E_\parallel = 10$ GeV to achieve the semiclassical domain, where the energy level density is high, as it is needed for quantum chaos investigations.

\begin{figure}[htbp]
\vspace*{-2mm}\centering
\includegraphics[width=\textwidth]{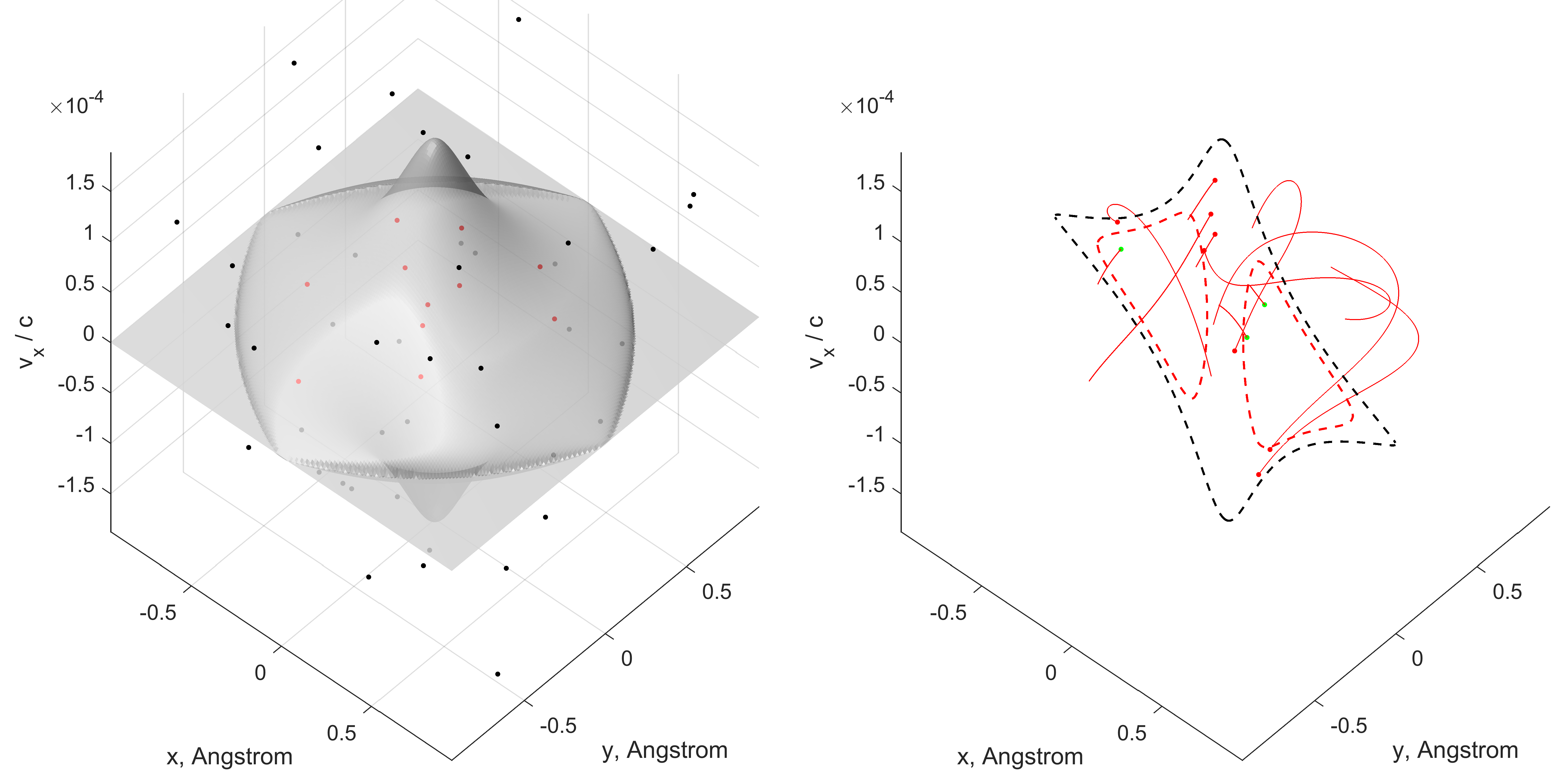}  
\vspace*{-7mm}\caption{\label{method} \emph{Left:} the surface $E = H(x,y,v_x,v_y=0)$ that bounds the phase space domain (\ref{permitted.domain}) permitted for motion under the laws of classical mechanics. The random points in this domain are plotted in red. \emph{Right:} the phase trajectories started from these points are traced up to their intersection with the plane $y = 0$. If the final point falls into a regular motion domain on the Poincar\'e section, the initial point also belongs to the regular motion domain in the phase space.}
\end{figure}

The semiclassical mean level density is described by the integral \cite{Berry} in the 4-dimensional phase space with the classical counterpart $H(x,y,p_x , p_y) $ of the quantum Hamiltonian (\ref{Hamiltonian}):
\begin{equation}\label{density.via.Berry.Robnik.general}
\rho (E_\perp) = (2\pi\hbar)^{-2} \int dx dy dp_x dp_y \, \delta \left( E_\perp - H(x,y,p_x , p_y) \right) =
\end{equation}
\begin{equation}\label{density.via.Berry.Robnik.general.1}
= \frac{2}{(2\pi\hbar)^2} \int  \frac{dx  dy  dp_x}{ |v_y (x,y,p_x) |}  \,,
\end{equation}
where $v_y$ is the $y$-component of the electron's velocity, and the integration in (\ref{density.via.Berry.Robnik.general.1}) is performed over the domain 
\begin{equation}\label{permitted.domain}
\frac{c^2p_x^2}{2E_\parallel} + U(x,y) \leq E_\perp \,.
\end{equation}
The integral (\ref{density.via.Berry.Robnik.general.1}) is computed using Monte-Carlo method. Under that, if the random point falls into a regular motion domain (see figure \ref{method}), its contribution is accounted both in the total density of states and in the density of states related to the regular motion domain. The boundaries of the regular motion domains are found using Poincar\' e sections method (see, e.g., \cite{AhSh}).

\section{Results and discussion}

\begin{figure}[htbp]
\vspace*{-2mm}\centering
\begin{minipage}[b]{10pc}
\includegraphics[width=\textwidth]{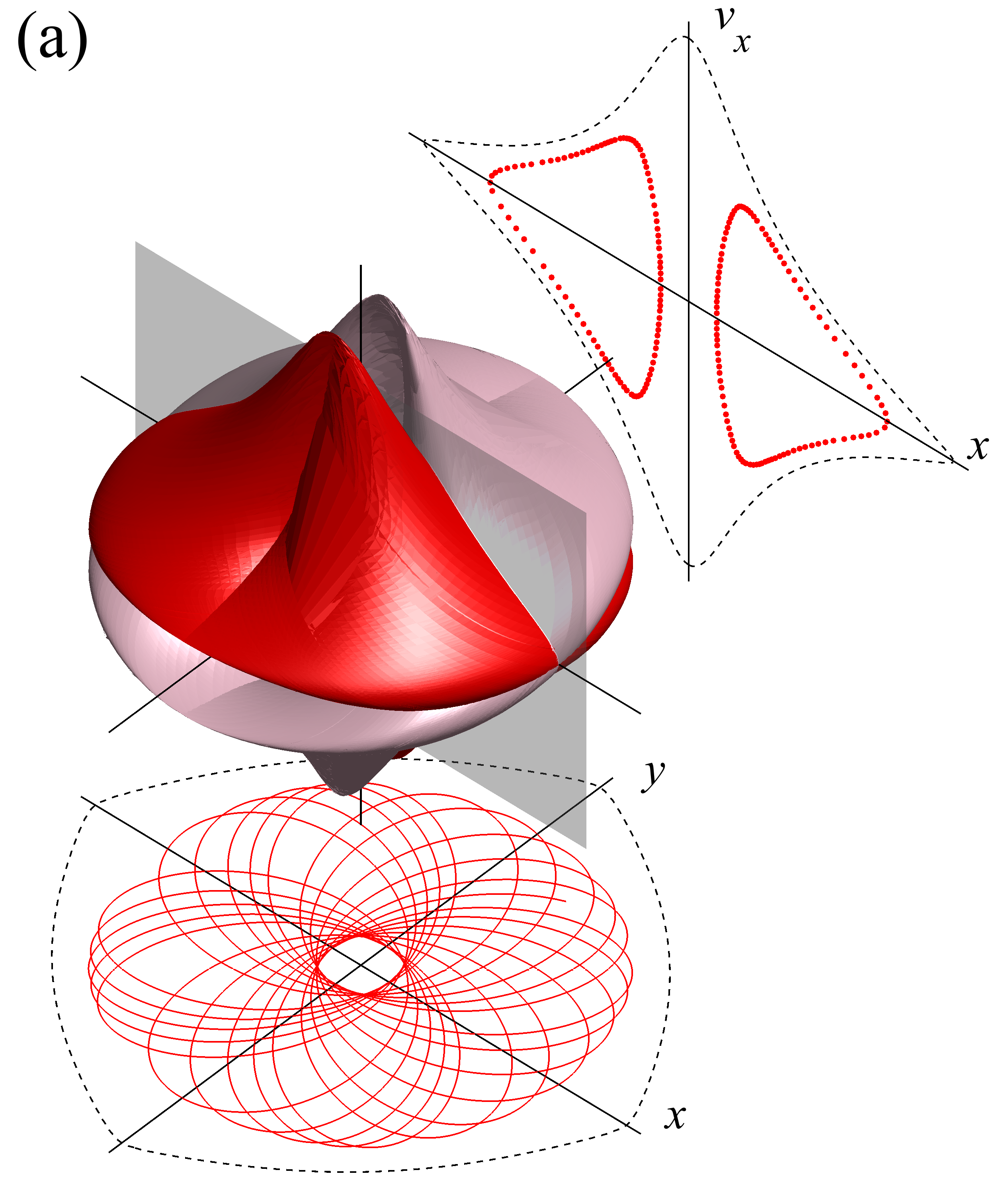} 
\end{minipage}
\begin{minipage}[b]{14pc}
\includegraphics[width=\textwidth]{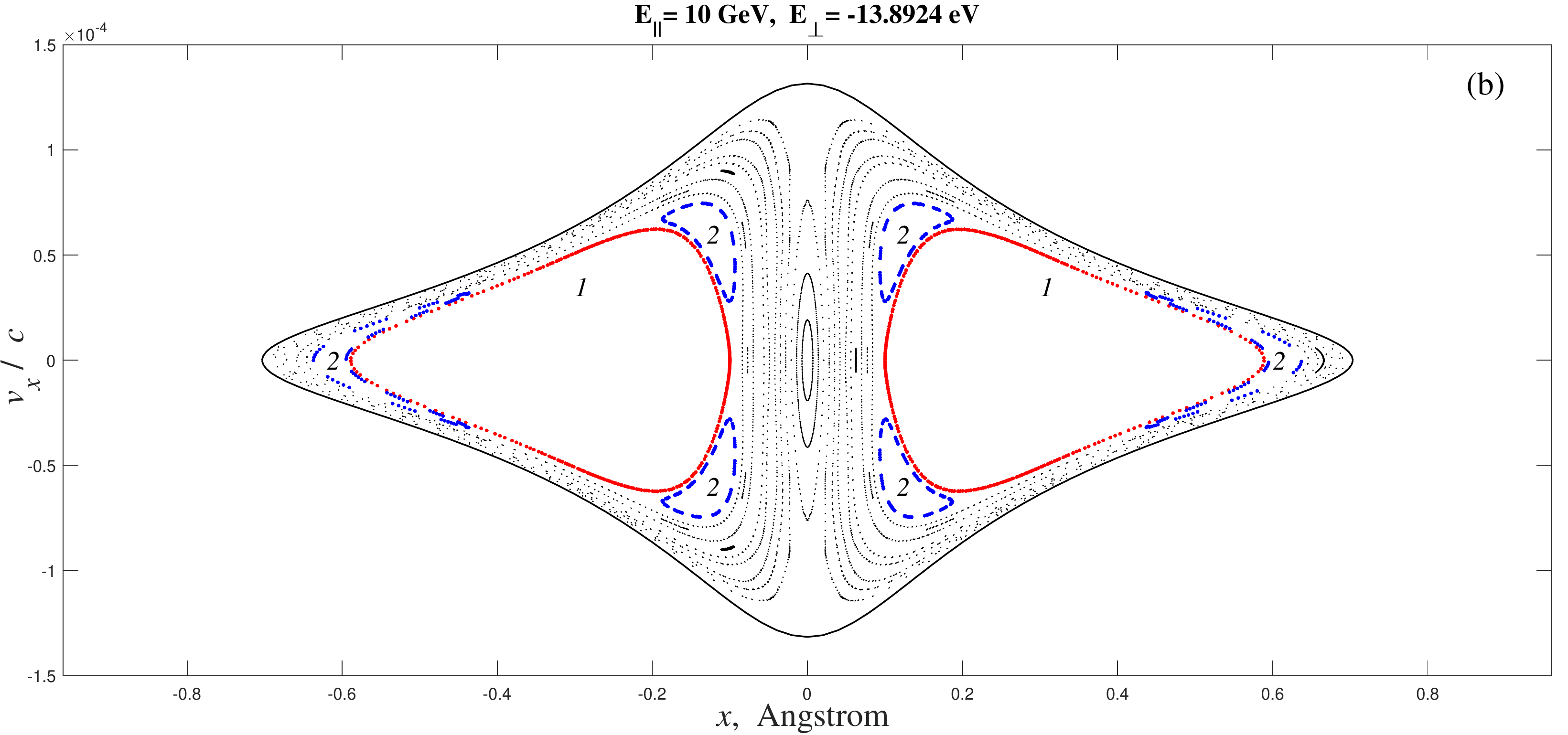}  \\
\includegraphics[width=\textwidth]{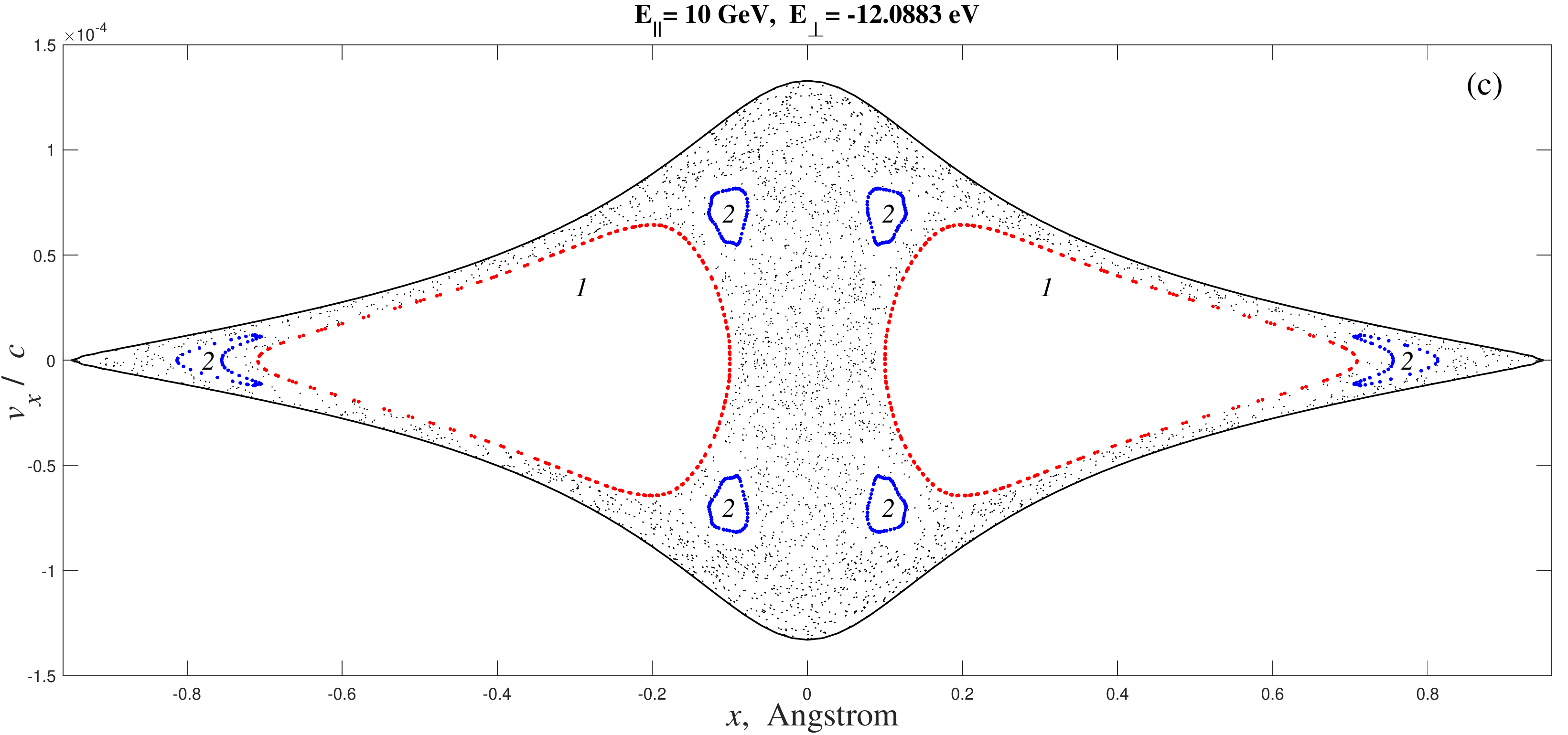}
\end{minipage}
\begin{minipage}[b]{10pc}
\includegraphics[width=\textwidth]{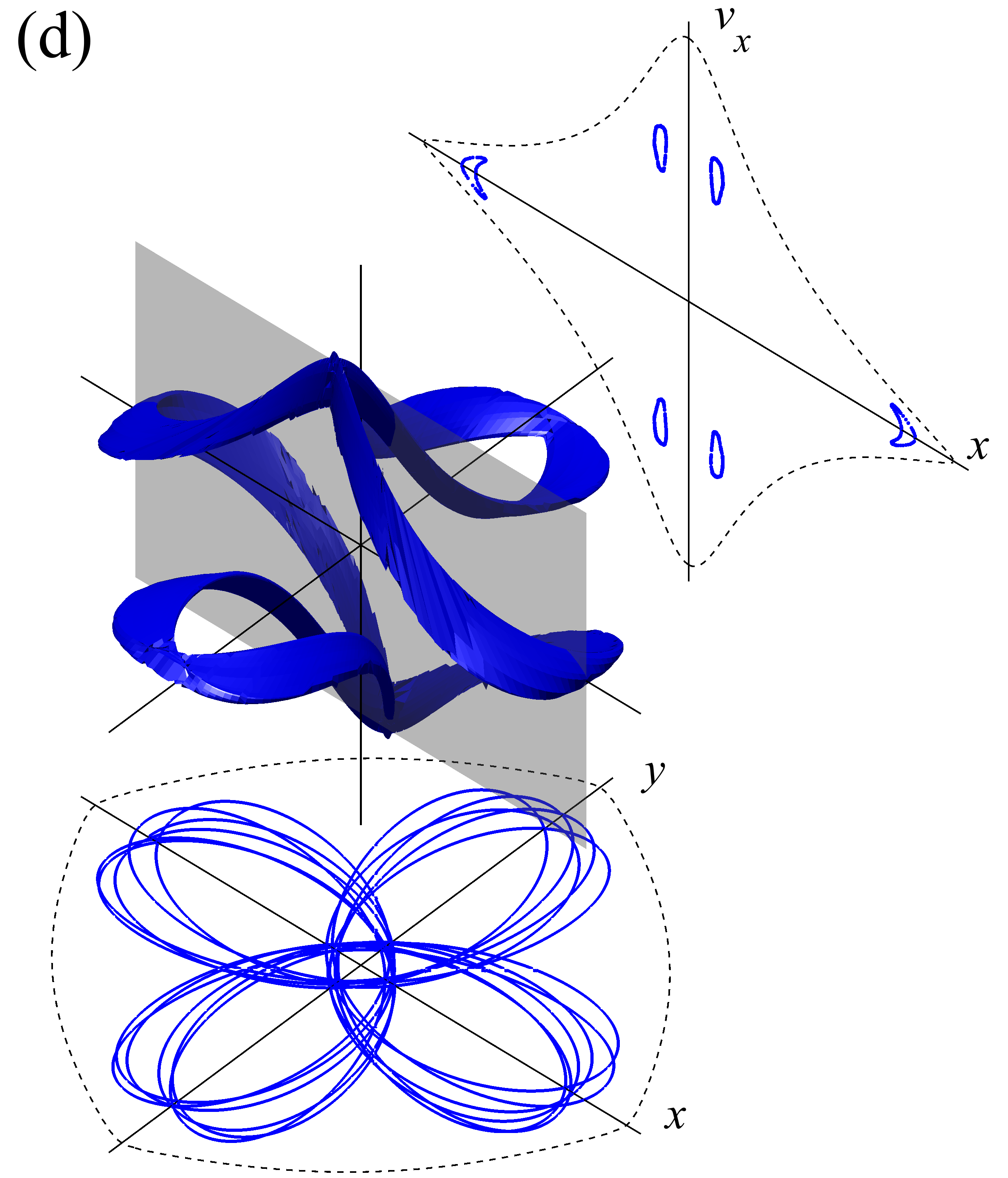} 
\end{minipage}
\vspace*{-2mm}\caption{\label{Poincar\'e.el} \emph{Middle:} Poincar\'e sections for the lowest and highest transverse energy levels of the electron with the energy of  longitudinal motion 10 GeV in the range under consideration. \emph{Left and right:} projections of the phase space domains of types 1 (left) and 2 (right) onto the 3-dimensional space $(x, y, v_x)$ as well as the corresponding orbits and Poincar\'e sections.}
\end{figure}

So, the electrons channeling along $[100]$ direction move in a
weakly disturbed, almost axially symmetric potential (\ref{U.el}). It leads to the approximate conservation of the electron's angular momentum (and hence to the regular dynamics) in the range far from the edges of the potential well.  As a consequence, there exists (among other regular domains) the regular motion domain marked \emph{1} on the Poincar\'e sections in the figure \ref{Poincar\'e.el} (b) and (c); the electron's transverse motion there is similar to ones in a central field (the orbit is presented on the bottom of the figure \ref{Poincar\'e.el} (a)). Such domain exists in the whole range of transverse motion energies under consideration from $E_\perp \approx -14$ eV (completely regular motion) up to $E_\perp = -12,0883$ eV (the upper edge of the potential well).

\begin{figure}[htbp]
\vspace*{-2mm}\centering
\includegraphics[width=0.49\textwidth]{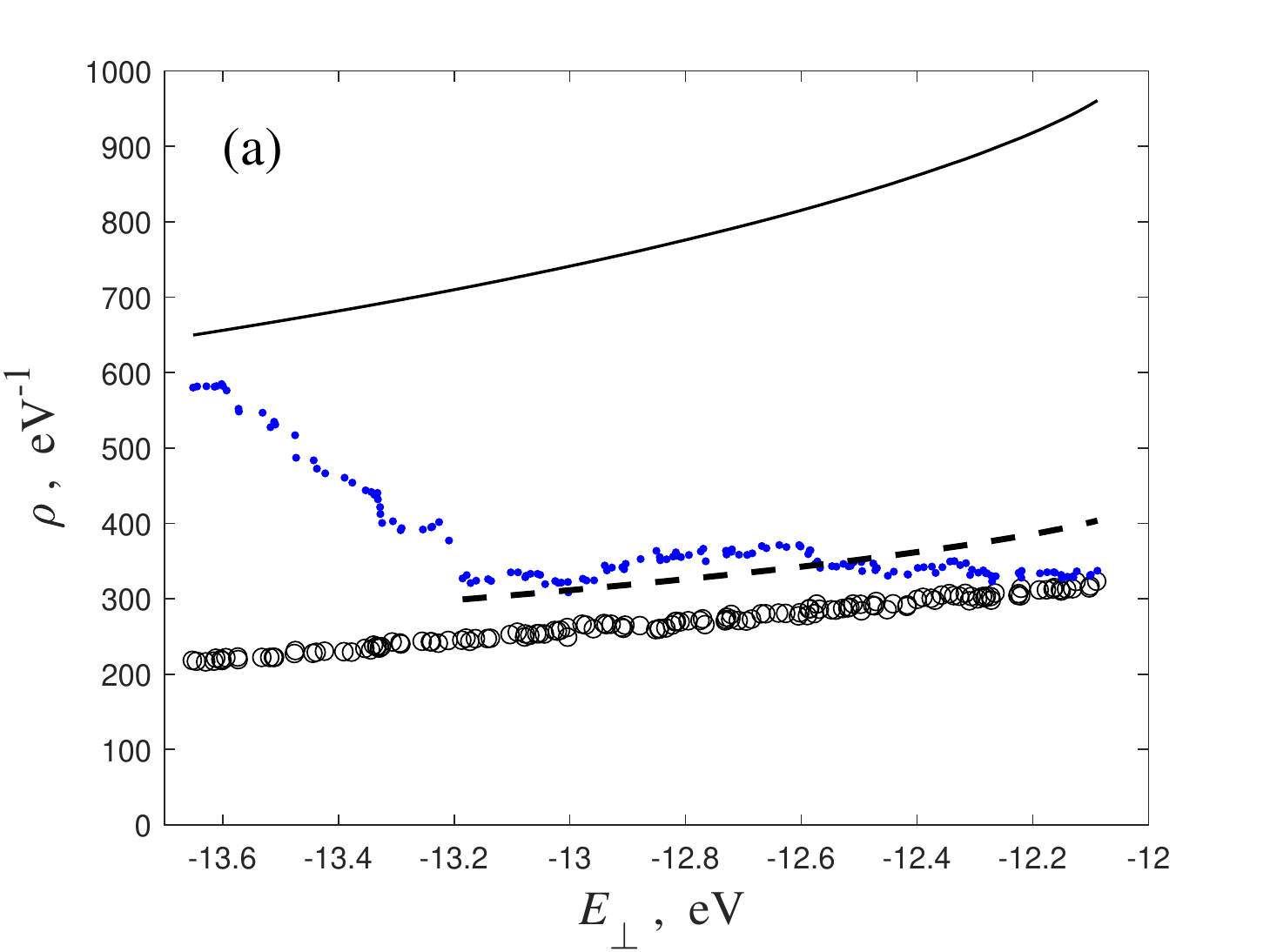} 
\includegraphics[width=0.49\textwidth]{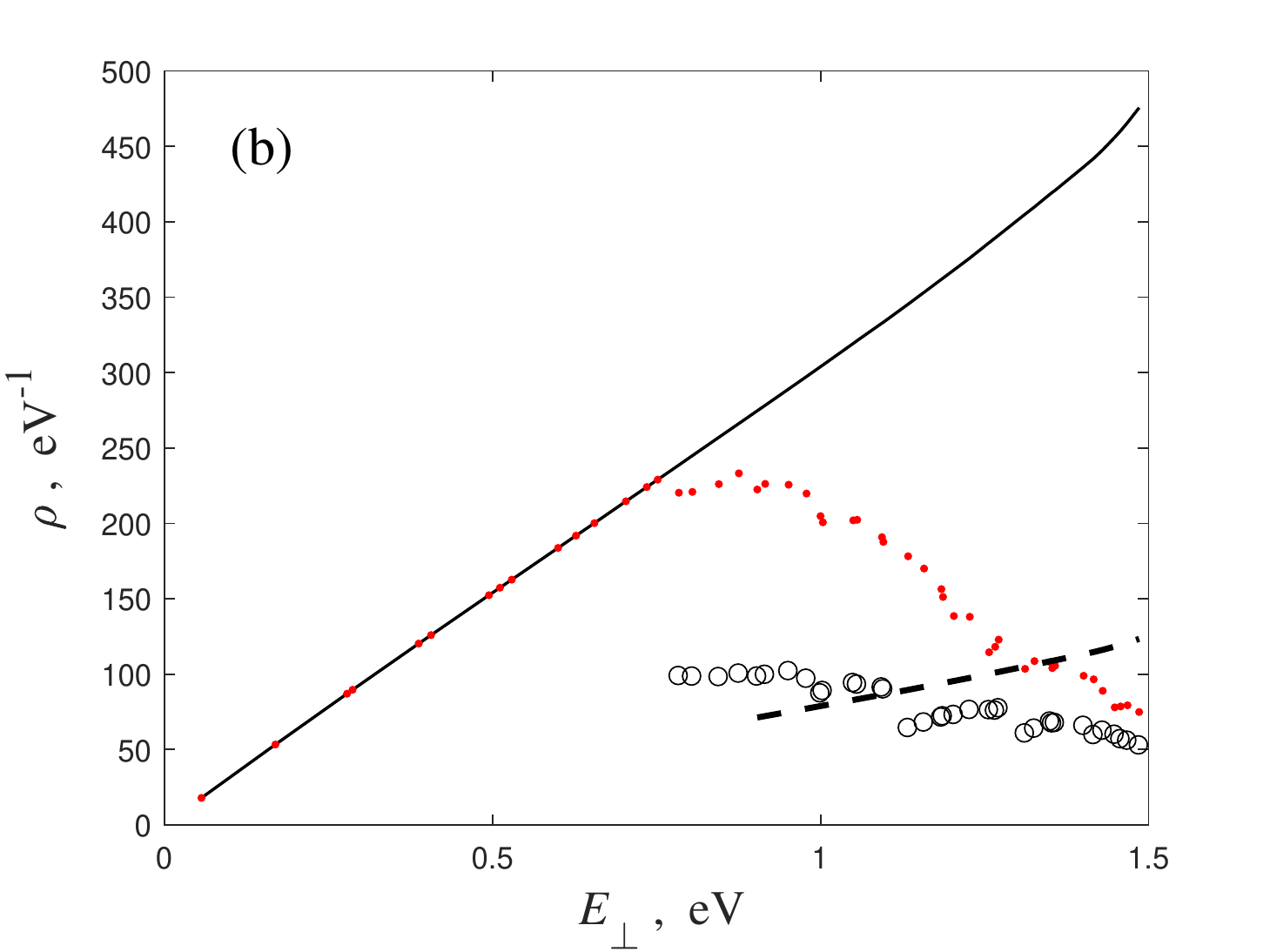} 
\vspace*{-2mm}\caption{\label{density} (a)  Semiclassical mean energy level density (\ref{density.via.Berry.Robnik.general.1}) (solid line), the contribution to it from the domain of the type 1 (circles), and the total contribution from all domains of regular motion (dots; errors are due to difficulty of precise determination of the domains' boundaries). The dashed line corresponds to the average value of the regular motion contribution in the interval under consideration that is equal to 42\%. (b) The same for the channeling positrons, while the circles are referred to the contribution from the regular motion domains of the types 1, 2 and 3 in figure \ref{Poincar_pos} (or the domain enveloping them for the lower energy levels).}
\end{figure}

It turns out that the relative contribution of this domain to the mean level density $\rho_1/\rho$ is approximately constant in the energy range under consideration, $-13.19\leq E_\perp \leq -12.0883$ eV; it is about 34\%. The average contribution from all regular motion domains in that range is about 42\% (dashed line in the figure \ref{density} (a)).

The mean level density $\rho$ (\ref{density.via.Berry.Robnik.general.1}) as well as the contributions to it from the regular motion domains are presented in figure \ref{density} (a). Note that $\rho$ varies in the range under consideration. In this case the generic set of energy levels has to be subjected the so-called unfolding procedure \cite{Reichl}, which leads to dimensionless values of $s$ with the mean levels density $\rho =1$ for the $E_\perp$ range under consideration.

\begin{figure}[htbp]
\vspace*{-1mm}\centering
\includegraphics[width=0.325\textwidth]{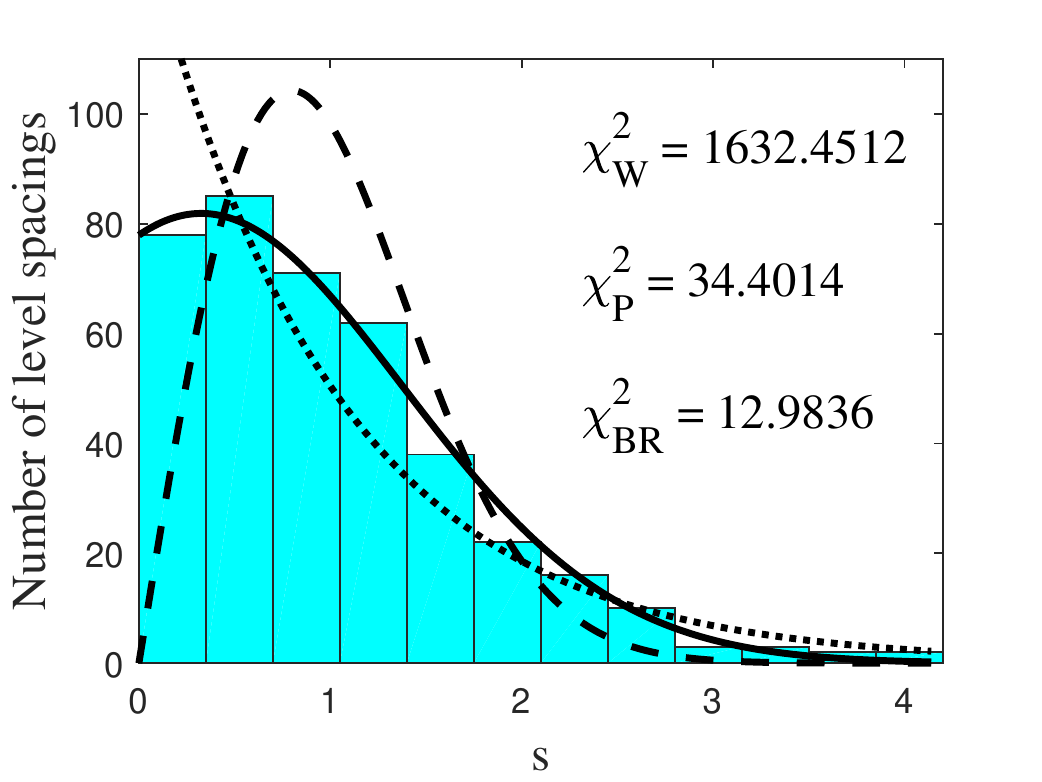} 
\includegraphics[width=0.325\textwidth]{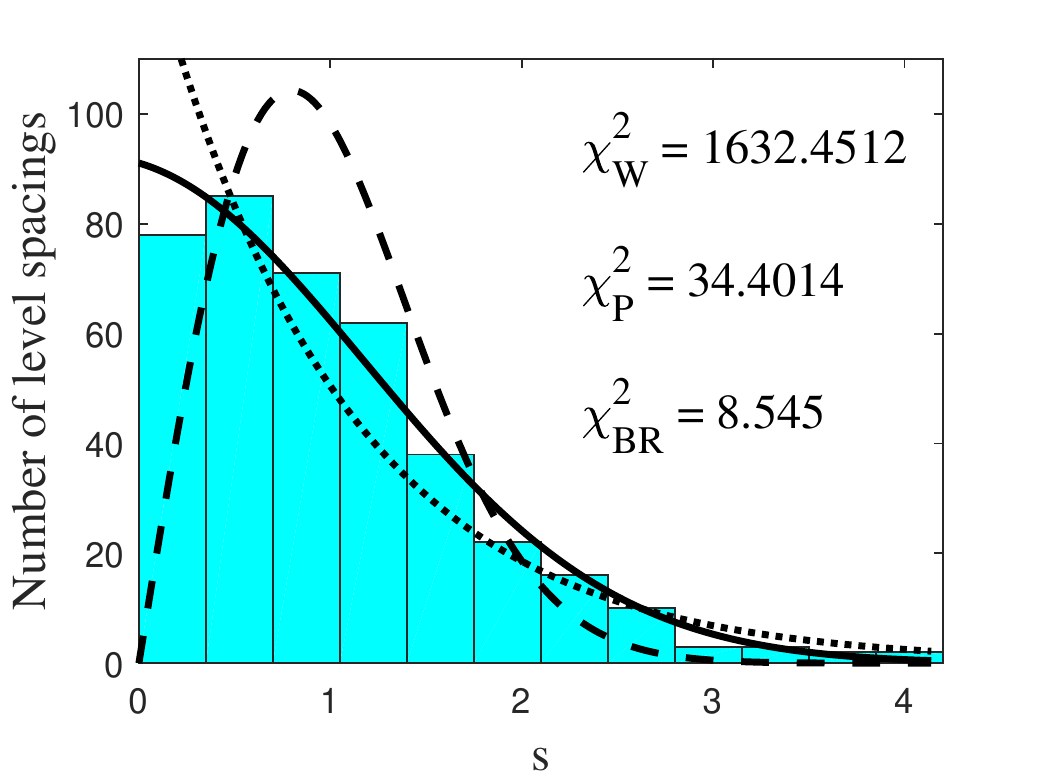} 
\includegraphics[width=0.325\textwidth]{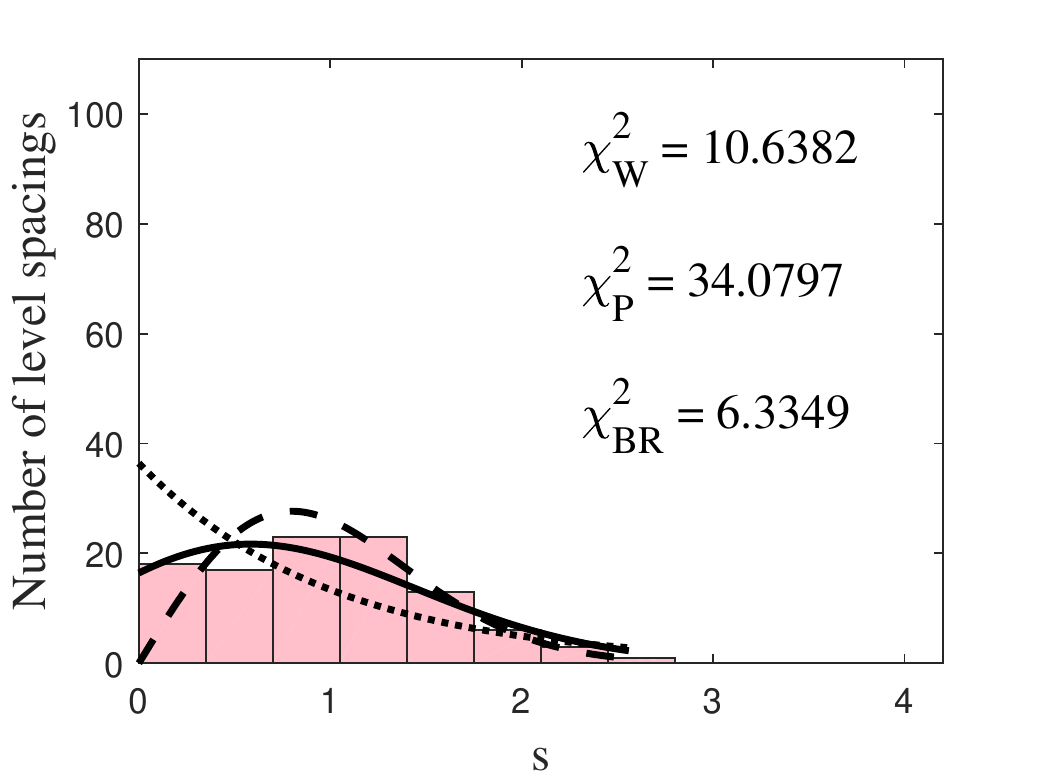}
\vspace*{-2mm}\caption{\label{histo} \emph{Left:} nearest-neighbor spacing distribution for the $E_\parallel = 10$ GeV channeling  electrons in the $-13.19\leq E_\perp \leq -12.0885$ eV range in comparison with Wigner (\ref{Wigner}), Poisson (\ref{Poisson}), and Berry--Robnik (\ref{Berry.Rob}) distributions (dashed, dotted, and solid lines, respectively); $\rho_1 = 0.34$. \emph{Middle:} the same for  $\rho_1 = 0.42$. \emph{Right:} the same for the channeling positrons in the $0.9\leq E_\perp \leq 1.4886$ eV range with $\rho_1 = 0.26$.}
\end{figure}

Note also that the channeling particle's eigenstates of the transverse motion can be classified using the group theory. The
potential (\ref{U.el}) possesses the symmetry of the square thus isomorphic to dihedral group $D_4$ (or $C_{4v}$). This group has four one-dimensional
irreducible representations and one two-dimensional one, denoted
$A_1$, $A_2$, $B_1$, $B_2$, $E$ \cite{group, LL3}. So, the eigenstates of the first four types are non-degenerated while the eigenstates of the last type are twice degenerated. The histogram in the figure \ref{histo} represents the levels spacings distribution for that four types of non-degenerated levels. 

The levels spacings distribution is compared to the predictions (\ref{Wigner}), (\ref{Poisson}) and (\ref{Berry.Rob}), the last case with the minimal and the average estimations of the regular motion domains contribution, $\rho_1/\rho\approx 34$\% and $\rho_1/\rho\approx 42$\% (left and middle panels in the figure \ref{histo}, respectively). We see that Berry--Robnik distribution (\ref{Berry.Rob}) describes the level spacing distribution much better than pure Wigner or Poisson ones (that is confirmed by the $\chi^2$ values calculated for all three hypotheses). 

\begin{figure}[htbp]
\vspace*{-2mm}\centering
\includegraphics[width=\textwidth]{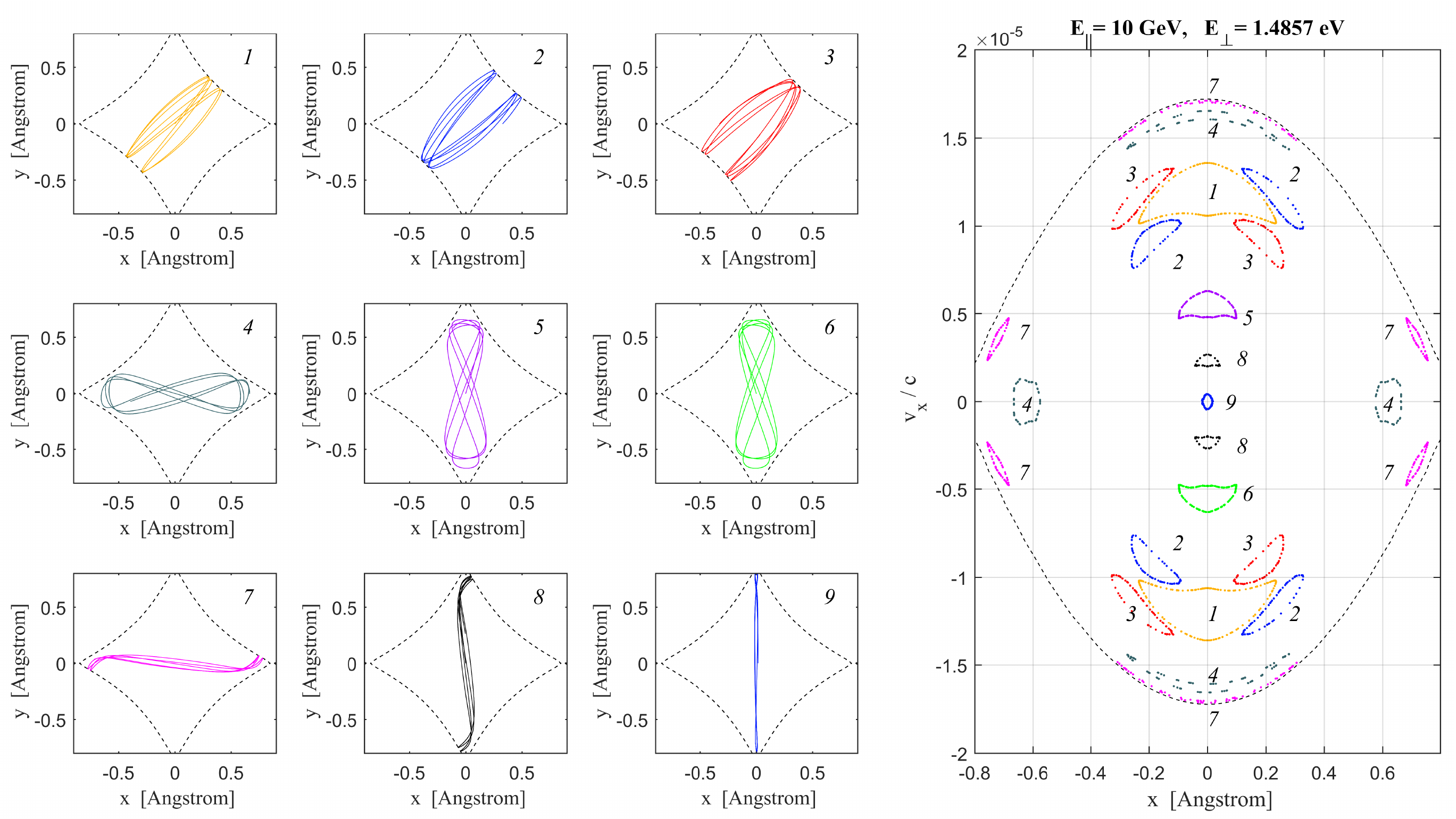} 
\vspace*{-10mm}\caption{\label{Poincar_pos} Main types of regular orbits and corresponding Poincar\'e sections for the channeling $E_\parallel = 10$ GeV positrons with $E_\perp = 1.4857$ eV.}
\end{figure}
\begin{figure}[htbp]
\vspace*{-1mm}\centering
\includegraphics[width=0.24\textwidth]{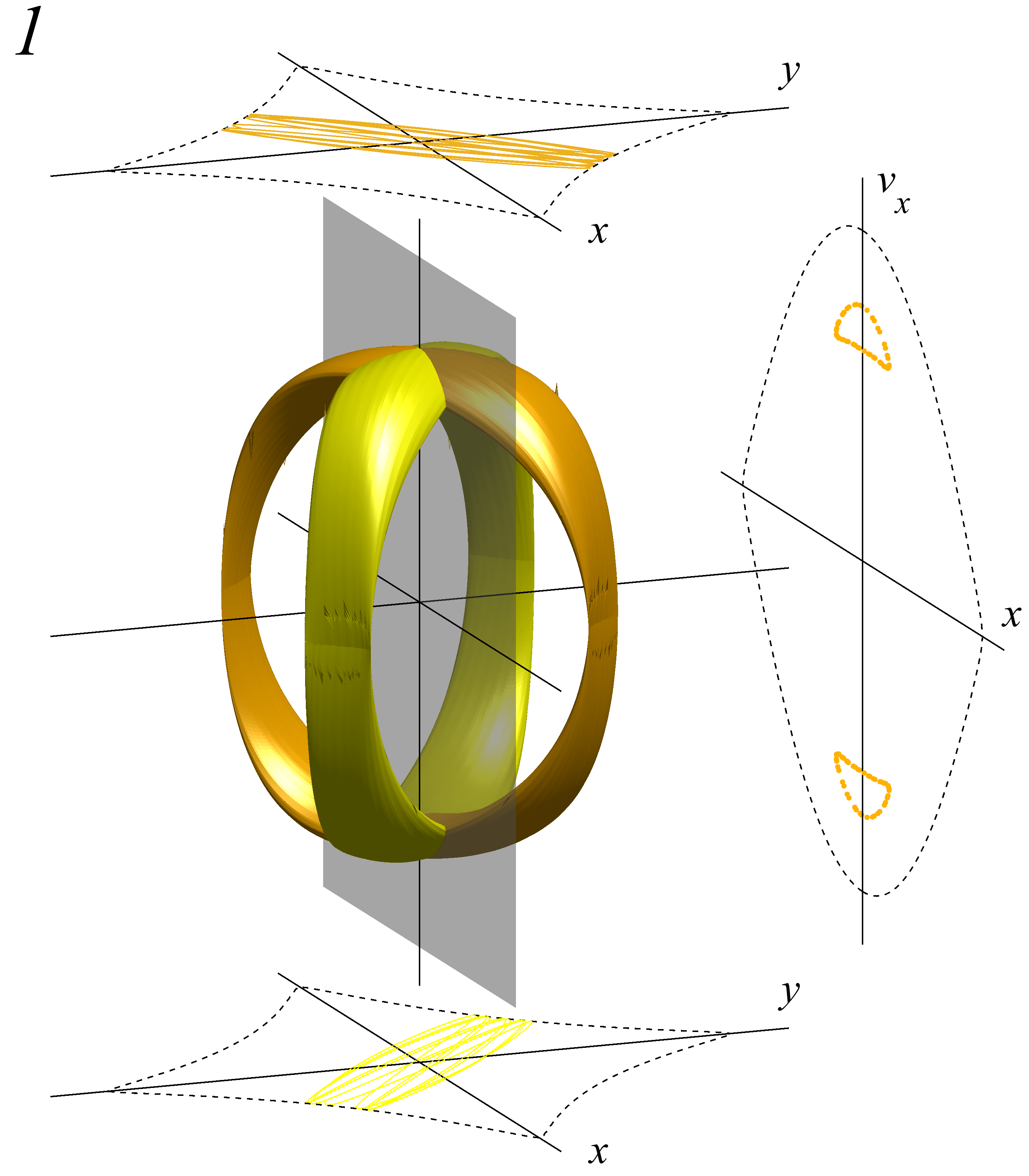} 
\includegraphics[width=0.24\textwidth]{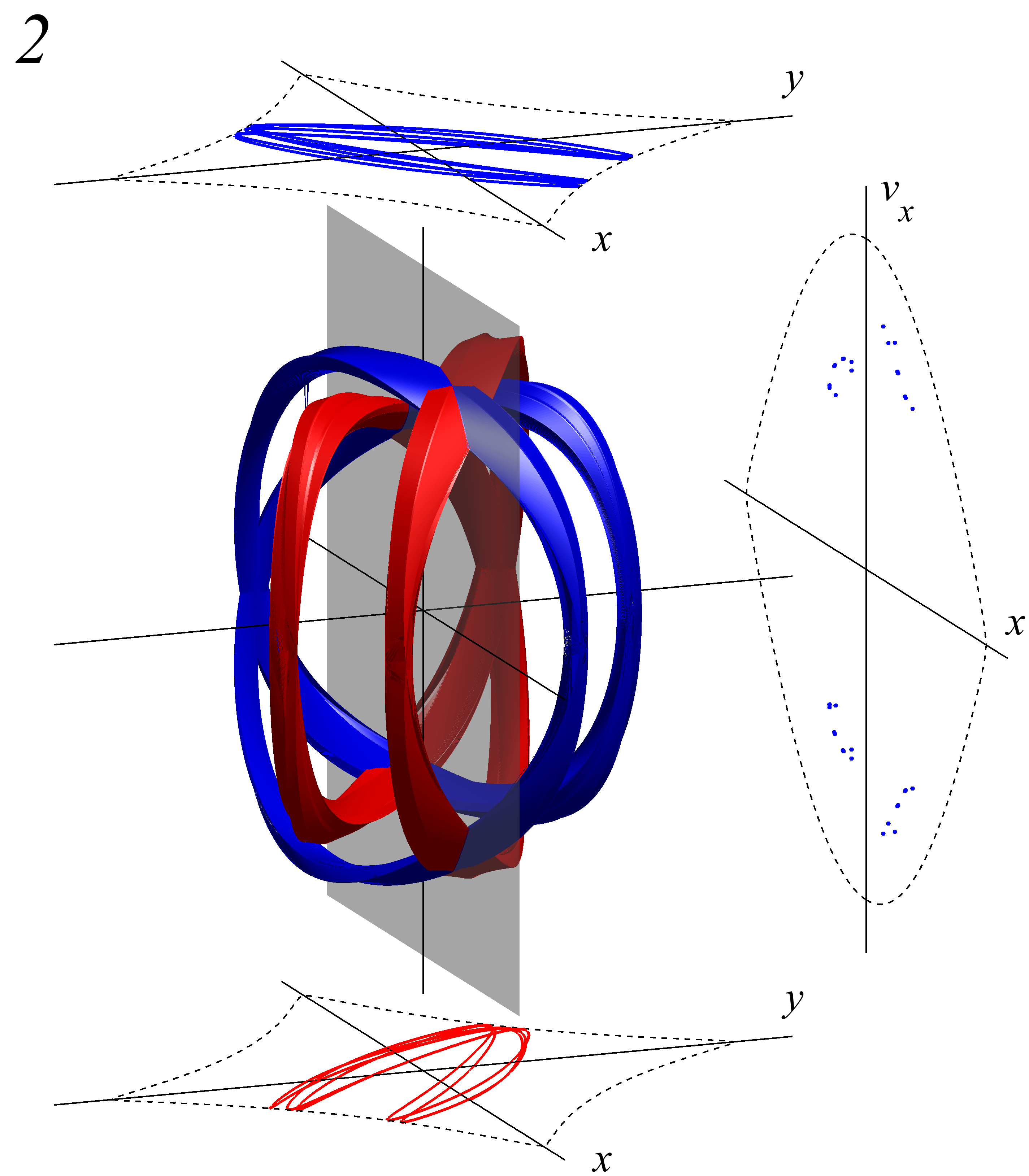}
\includegraphics[width=0.24\textwidth]{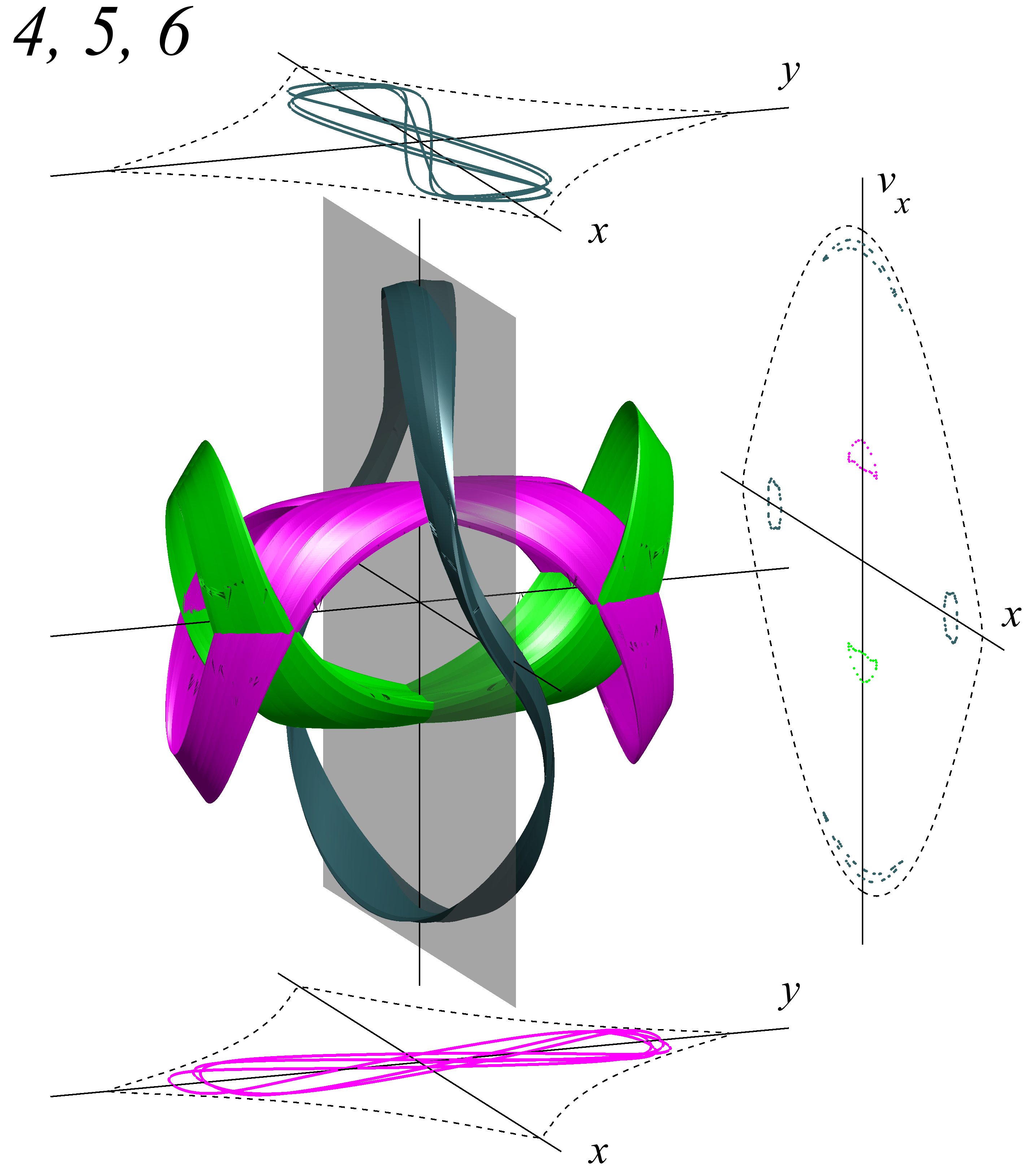} 
\includegraphics[width=0.24\textwidth]{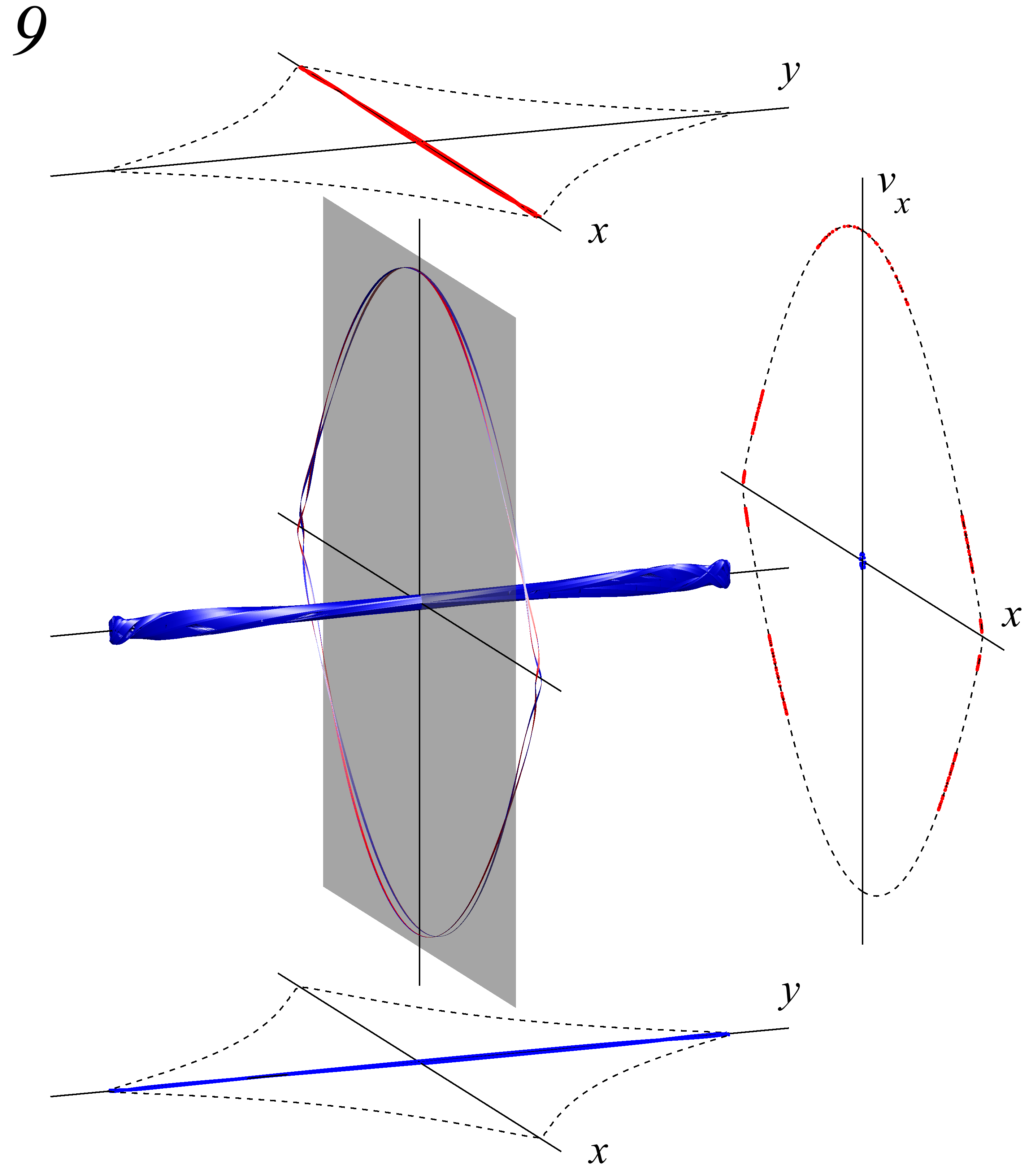}
\vspace*{-2mm}\caption{\label{domains_pos} The positron's regular motion domains of various sorts from the figure \ref{Poincar_pos} in the $(x, y, v_x)$ space. The panel \emph{1} presents the orbit \emph{1} from the figure \ref{Poincar_pos} (dark yellow) and its counterpart rotated on $90^\circ$ (bright yellow) making the same contribution to the total level density,  the panel \emph{2} presents the orbit \emph{2} (blue) and its counterpart (red); in these cases both counterparts make the same trace on the Poincar\'e section plane. The panel \emph{9} presents the orbit \emph{9} (blue) and its counterpart (red); in this case the Poincar\'e sections of these counterparts are essentially different. The panel \emph{4, 5, 6} presents the orbit \emph{4} (dark green) but not its counterpart and also the orbits \emph{5} and \emph{6} that differ from each other only by the direction of motion, clockwise or counterclockwise, their Poincar\'e sections are symmetric to each other (purple and bright green).}
\end{figure}

In contrast to the case of electron, the potential well for the channeling positrons (\ref{U.pos}) could not be considered as a slightly perturbed axially symmetric
well. Hence the structure of regular motion domains stays rather complex yet near the upper edge of the potential pit (see figures \ref{Poincar_pos} and \ref{domains_pos}).

There is no the energy range with approximately constant relative contribution of the regular motion domains to the mean level density $\rho_1/\rho$. However we consider the energy range $0.9\leq E_\perp \leq 1.4886$ eV with the value $\rho_1/\rho$ about 26\% in it (dashed line in the figure \ref{density} (b)). We see that in this case Berry--Robnik function also describes the level spacing distribution better than Wigner or Poisson ones (figure \ref{histo}, right panel).

\section{Conclusion}

The energy levels of the transverse motion of $E_\parallel = 10$ GeV electrons and positrons channeling in the [100] direction of a silicon crystal are found using the spectral method of numerical integration of the time-dependent Schr\"odinger equation. The case of channeling in this direction is interesting for the quantum chaos investigations due to the co-existence of the regular and chaotic motion domains with given transverse motion energy value. The Berry-Robnik theory predicts the formula for the level spacing distribution for such cases.

We have estimated the relative contribution of the regular motion domains in the phase space in the semiclassical mean level density. This value is needed as a parameter in Berry-Robnik distribution.

There exists the electron's transverse energy interval where the contribution of the regular motion domains is approximately constant, satisfying the condition under which Berry--Robnik theory is built; it comprises approximately 42\%. The level spacing distribution in this interval is much better described by Berry-Robnik distribution rather than pure Wigner or Poisson ones ($\chi^2 = 8.545$ for 11 degrees of freedom that corresponds to $p = 0.66$; for Wigner and Poisson distributions we have $p$ values 0 and $3.1\cdot 10^{-4}$, respectively). 

There is no wide enough interval with approximately constant regular contribution in the case of channeling positrons. However, even in this case Berry-Robnik distribution, with the mean regular contribution as the parameter, describes the level spacing distribution better than Wigner and Poisson ones  ($\chi^2 = 6.3349$ for 11 degrees of freedom that corresponds to $p = 0.85$; for Wigner and Poisson distributions we have $p$ values 0.47 and  $3.5\cdot 10^{-4}$, respectively).

\end{document}